# The self-consistent charge density functional tight-binding (DFTB) theory study of carbon adatoms using tuned Hubbard U parameters

Jia Wang,[1,2,a)] Xing Dai,[1,2,a)] Wanrun Jiang,[1,2] Tianrong Yu,[1,2] Zhigang Wang[1,2,b)]

[1]*Institute of Atomic and Molecular Physics, Jilin University, Changchun, 130012, China*

[2]*Jilin Provincial Key Laboratory of Applied Atomic and Molecular Spectroscopy (Jilin University), Changchun, 130012, China*

The self-consistent charge density functional tight-binding (DFTB) theory is a useful tool for realizing the electronic structures of large molecular complex systems. In this study, we analyze the electronic structure of $C_{61}$, formed by fullerene $C_{60}$ with a carbon adatom, using the fully localized limit and pseudo self-interaction correction methods of DFTB to adjust the Hubbard U parameter (DFTB+U). The results show that both the methods used to adjust U can significantly reduce the molecular orbital energy of occupied states localized on the defect carbon atom and improve the gap between highest occupied molecular orbital(HOMO) and lowest unoccupied molecular orbital(LUMO) of $C_{61}$. This work will provide a methodological reference point for future DFTB calculations of the electronic structures of carbon materials.

**I. INTRODUCTION**

Carbon adatoms are one of the sources of intrinsic magnetism in low-dimensional carbon systems[1-3] such as graphene, carbon nanotubes (CNTs) and fullerenes. This kind of defect is usually a carbon atom absorbed at a C–C bond between two hexagons (bridge-sites) in a carbon system. Here, we explain the electron behavior of this kind of absorbed atom using a simple theoretical model, as shown in Fig. 1. In general, in this type of carbon system, every carbon atom has four valence electrons, three $sp^2$ electrons and one $p_z$ electron. However, two $sp^2$ electrons of the adatom form two sigma bonds with two $sp^2$ electrons from adjacent carbon atoms, leaving the $sp^2$ electron in the Y-direction (Fig. 1(a)). Meanwhile, the remaining $p_z$ electron of the adatom cannot bond with any electron from the other atoms (Fig. 1(b)). Therefore, there are two electrons that are spin-polarized for the adatom. Experimental studies have shown that these adsorbed defects have magnetic moments which arise from the spin polarization of adatomic electrons.[4] Several previous studies using conventional density functional theory (DFT) have also confirmed this theoretical model.[3, 5-8]

Compared to first-principles, in particular DFT, the self-consistent charge density functional tight binding method

---

[a)] These authors contributed equally to this work
[b)] Electronic mail: wangzg@jlu.edu.cn

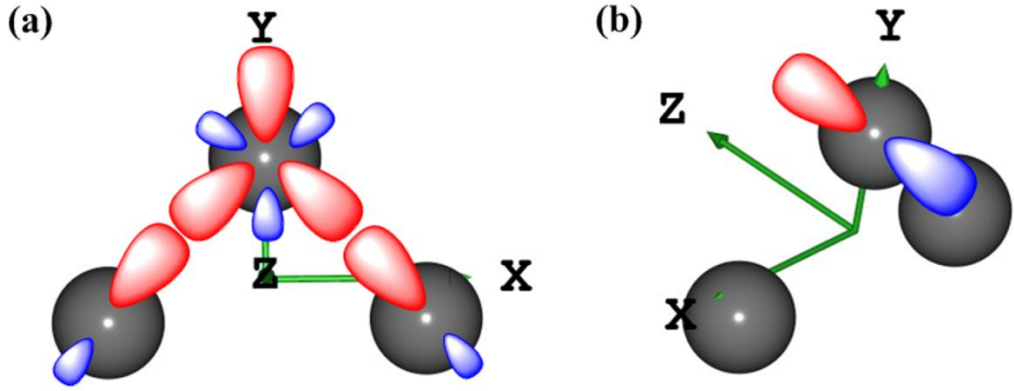

FIG. 1. Electronic structure of a typical carbon adatom defect. (a) The $sp^2$ electrons in the Y-direction of the adatom. (b) The remaining $p_z$ electrons of the adatom.

(SCC-DFTB, hereafter abbreviated as DFTB) provides a new extension and analysis platform for quantum calculations[9] and is aimed at giving a good approximation for DFT results. In DFTB, only the two-center integral Hamiltonian and overlap matrix elements are included, with the three-center integral neglected and a short-range repulsive potential used[10]. Thus, the DFTB method has attracted much attention because of its higher computational efficiency[11] and relatively reliable results. For both solid[12-15] and molecular systems[16], the original form (i.e. non-SCC form)[17] and the SCC form[18] of the DFTB method all show wide applications. However, we must be especially careful in investigating some particular areas (mainly d and f regions) of the periodic table by DFT and its related methods, owing to false DFT self-interactions in these systems[19]. Therefore, the introduction of the Hubbard U model into the original Hamiltonian is required, best known as DFT+U (i.e. LDA+U and GGA+U)[20, 21]. Of course, this kind of correction can also be introduced into the DFTB method. Theoretical studies show that the DFTB+U method can accurately calculate NiO molecules and the highest occupied molecular orbital-lowest unoccupied molecular orbital (HOMO-LUMO) gap is in good agreement with experimental values, when the Hubbard U parameter is adjusted.[22]

Molecular systems containing d or f electrons, such as Fe, Co and Ni, are widely studied due to their spin polarization which has practical applications[23, 24]. Previous studies indicate that DFT+U can change their electronic structure and make their properties, such as energy gaps, magnetic moments and charge, conform to experimental values.[25, 26] High-precision studies show that p electrons also contain intrinsic magnetism, especially in low-dimensional carbon materials.[27-29] Recently, significant attention has been paid to the behavior of localized p electrons[30] and previous experimental reports of spin polarization phenomenon of typical carbon materials have urged for the further development of theoretical studies in this important area. It has been found that charge transfer between carbon and metal atoms adsorbed on the top site is strongly influenced by the U parameter, and the spin polarization induced by chemisorptions also depends on the choice of U parameter for graphene[31]. The DFT+U method has also been used to investigate spin polarization formation in fullerene $C_{20}$.[32] For the structure of graphene with hydrogen atom adsorption, it was discovered that the stability of the ferromagnetic state is very sensitive to the U parameter used.[33] For the structure of graphene nanoflakes, research shows that the nature of the magnetic order strongly depends on U[34]. As a result, the method for Hubbard U parameter modulation has a non-negligible contribution to spin polarization in carbon materials. In addition, the DFTB method is suitable for handling large-scale systems and short period elements, such as C, H, O and N. It is hopeful to get the electronic properties of larger systems on the nanoscale based on DFTB method, particularly for low-dimensional carbon materials. Therefore, exploring DFTB+U is important for both research and application.

## II. COMPUTATIONAL DETAILS

It is considered that the [6, 6] bond (connecting two hexagons) is the most common bond in fullerene, CNT and graphene, so the typical adatom-defective $C_{61}$ with a carbon atom absorbed on the [6, 6] bond of fullerene $C_{60}$ is selected as the model in this work, as shown in Fig. 2(a) and Fig. (b). This model is consistent with adsorbed defects on the surfaces of graphene and CNTs, as previously reported.[35, 36]

We used the DFTB method[37, 38] to calculate the geometric and electronic structures of $C_{61}$. The simple DFTB method can be regarded as a special case for U = 0. We used the spin-polarized SCC-DFTB Hamiltonian and linear spin-polarized algorithm. The standard spin constants of PBE-GGA we used for all carbon atoms[39], Wss = -0.031, Wsp = -0.025, Wps = -0.02 and Wpp = -0.023 Ha (Wss is the constants with respect to s shells coupled, as same for Wsp, Wps, Wpp). The convergence criterions of optimization were set to $10^{-8}$ eV for SCC and $10^{-6}$ eV/Å for the force. At present, there are two forms for LDA+U correction[40] supported in the DFTB method, the fully localized limit (FLL) and the pseudo self-interaction correction (pSIC). According to previous studies, for solving the spin in SCC-DFTB, the contributions of the FLL and pSIC corrections[41, 42] to the potential are:

(FLL) $\quad \Delta V_{\mu\nu}^{\sigma} = -\alpha(U-J)_l^{atomic}\left(n_{\mu\nu}^{\sigma} - \frac{1}{2}\delta_{\mu\nu}\right)$ \hfill (1)

(pSIC) $\quad \Delta V_{\mu\nu}^{\sigma} = -\alpha\frac{(U-J)_l^{atomic}}{2}n_{\mu\nu}^{\sigma}$ \hfill (2)

Here, μ and ν are in the same l shell of an atom. The contributions of the two corrections to the total energy are:

$$\Delta E^{FLL} = -\alpha \sum_a^{atoms} \sum_{l \in a} \frac{(U-J)_l^{atomic}}{2} \sum_\sigma \sum_{\mu\nu} \left( (n_{\mu\nu}^\sigma)^2 - n_{\mu\nu}^\sigma \right)_{\nu\mu \in l} \qquad (3)$$

$$\Delta E^{pSIC} = -\alpha \sum_a^{atoms} \sum_{l \in a} \frac{(U-J)_l^{atomic}}{2} \sum_\sigma \sum_{\mu\nu} (n_{\mu\nu}^\sigma)_{\nu\mu \in l}^2 \qquad (4)$$

These potential energies can affect the energy of occupied states for a specified shell of an atom and usually enhance localization of these states. The FLL potential lowers the energy of occupied states, which are localized at specified atomic shells, and increases the energy of unoccupied states. The pSIC potential is capable of correcting the local part of self-interaction errors, so it is possible to reduce the energy of occupied states. Since carbon atoms may undergo hybridization for the 2s and 2p electrons, we considered the U correction not only for 2s shells, but also for 2s and 2p shells of the carbon atoms in this work. We performed a series of calculations with U ranging from 0.01 to 0.1 Ha. All calculations were carried out using the DFTB+ code.[43]

We mainly studied the effect of DFTB+U on geometric structures, molecular orbital energy order, charge and spin population and the HOMO-LUMO gap. Previous studies have shown that using the Hubbard U model can significantly improve the value of the HOMO-LUMO gap, making the value closer to experiment.[22] However, the HOMO-LUMO gap is not defined rigorously in linear spin-polarized calculations. In this work, the $HOMO_\alpha$-$LUMO_\beta$ gap is defined as the HOMO-LUMO gap, with $HOMO_\alpha$ and $LUMO_\beta$ corresponding to the highest energy occupied state and the lowest energy unoccupied state, respectively, according to DFT results.[44] This definition has also been used in other studies with spin-polarized calculations.[45]

## III. RESULT AND DISCUSSION

We optimized the geometric structure of $C_{61}$ using the DFTB method without any U correction (i. e. U = 0). As shown in Fig. 2, there are two spin-polarized electrons in $C_{61}$, which are localized on the adatom and are not paired or bonded with any others. The remaining MOs are doubly occupied. Meanwhile, the energy of MO(I) is higher than MO(II) by about 0.44 eV. However, a serious problem is that the energy of LUMO-β (MO(II')) is lower than HOMO-α (MO(I)), which results in the energy of the unoccupied state being lower than the occupied state, meaning the HOMO-LUMO gap is negative. In other words, the MO(I) energy of the state localized on the adatom is not low enough with a U = 0. However, this phenomenon, in principle, can be significantly improved by appropriate forms like the LDA+U correction. Therefore, we will calculate and discuss the electronic structure of $C_{61}$ using the DFTB+U method (i. e. U ≠ 0) in the following.

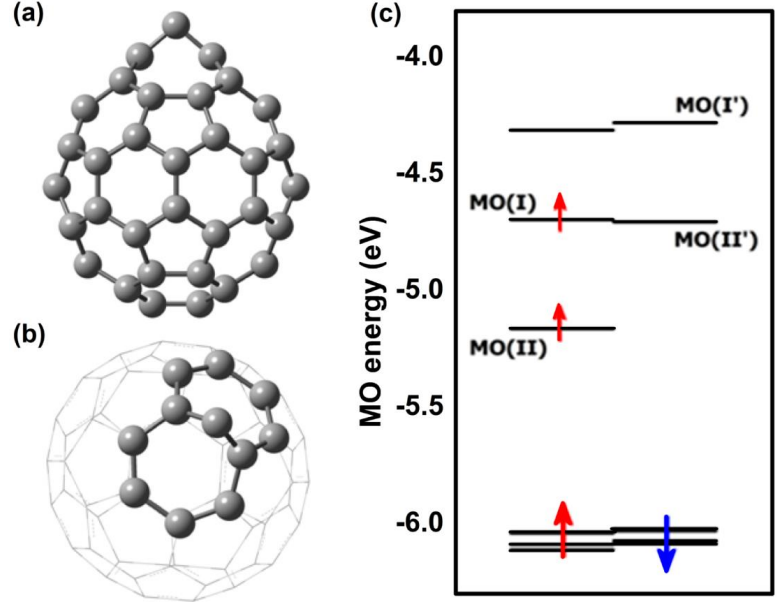

**FIG 2.** Geometric and electronic structures of $C_{61}$ calculated by the DFTB method. (a) Overall perspective, (b) partial perspective and (c) MO energy. Here, the energy difference between MO(I) and MO(II') is defined as the HOMO-LUMO gap.

First, we corrected the 2p shell of all carbon atoms using the FLL and performed re-optimization by the DFTB+U (FLL, p) method. The geometric parameters are summarized in Table I. We found that, with an increase in U, the length of the C–C bond remained almost unchanged, but the vertex angle of the adsorbed defect decreased to even less than 1°. Hence, the U(FLL, p) correction does not significantly affect the geometric results for $C_{61}$ compared to the DFTB method without correction. When the U (FLL, p) correction is considered, the charge of the adatom does not strongly change and the net spin distribution gradually increases with U in the range of 0.01–0.1.

TABLE I. Geometric parameters, charge, net spin distribution and HOMO-LUMO gap of $C_{61}$ calculated with the DFTB+U (FLL, p) method. We defined the C–C bond length between the adsorbed atom and adjacent carbon atoms as the symbol $l$, and the vertex angle of the adsorbed defect as $\angle A$.

| U(FLL, p) | $l$ | $\angle A$ | Adatom Charge | Adatom Spin | Gap (eV) |
|---|---|---|---|---|---|
| 0.00 | 1.43 | 99.53 | -0.34 | 1.29 | / |
| 0.01 | 1.43 | 99.42 | -0.34 | 1.32 | 0.10 |
| 0.02 | 1.43 | 99.32 | -0.34 | 1.34 | 0.22 |
| 0.03 | 1.43 | 99.23 | -0.35 | 1.35 | 0.33 |
| 0.04 | 1.44 | 99.15 | -0.35 | 1.37 | 0.45 |
| 0.05 | 1.44 | 99.08 | -0.35 | 1.39 | 0.57 |
| 0.06 | 1.44 | 99.01 | -0.35 | 1.39 | 0.59 |
| 0.07 | 1.44 | 98.95 | -0.35 | 1.41 | 0.62 |
| 0.08 | 1.44 | 98.90 | -0.35 | 1.41 | 0.64 |
| 0.09 | 1.44 | 98.85 | -0.35 | 1.42 | 0.66 |
| 0.10 | 1.44 | 98.80 | -0.35 | 1.43 | 0.69 |

The calculated MO energy of $C_{61}$ is shown in Fig. 3. Compared with U = 0, the non-zero U correction can gradually lower the energy of MO(I), so that the energy of MO(I) becomes less than that of LUMO-$\beta$. The decrease in energy for such a localized occupied state led to an increase in the net spin distribution of adatoms, as shown in Table I. We found that the (FLL, p) correction can slightly increase the orbital energy of LUMO-$\beta$, however, it has only a small effect on MO(II), empty orbitals of higher energy and approximately doubly occupied orbitals of lower energy. Therefore, by including the U correction, we can achieve the effect of separating the occupied and unoccupied states. We observed from Fig. 3 that the orbital energy of MO(I) gradually reduces with increasing of U. When U is larger than 0.06 Ha, the orbital energy of MO(I) is even lower than that of MO(II). We linear fit the orbital energies of MO(I) and MO(II) and found that these two energies are equal when the U (FLL, p) is 0.0501 Ha (1.36 eV), seen Fig. S1 in supplemental material[46]. The data in Table II show that the HOMO-LUMO gap increases strongly with increasing of U when the value of U is less than 0.05 Ha. The gap only increases marginally when U is larger than 0.05 Ha, which arise from the orbital energy of MO(I) has been lower than that of MO(II) and then the HOMO-LUMO gap is the energy difference between MO(II) and LUMO-$\beta$. Thus, U should be smaller than 0.05 Ha. From Fig. 3, we known that U has almost no impact on MO(II) and the LUMO-$\beta$ energy is not sensitive to the variation of U. As a result, the DFTB+U (FLL, p) method can reasonably improve the electronic structure and energy gap of $C_{61}$.

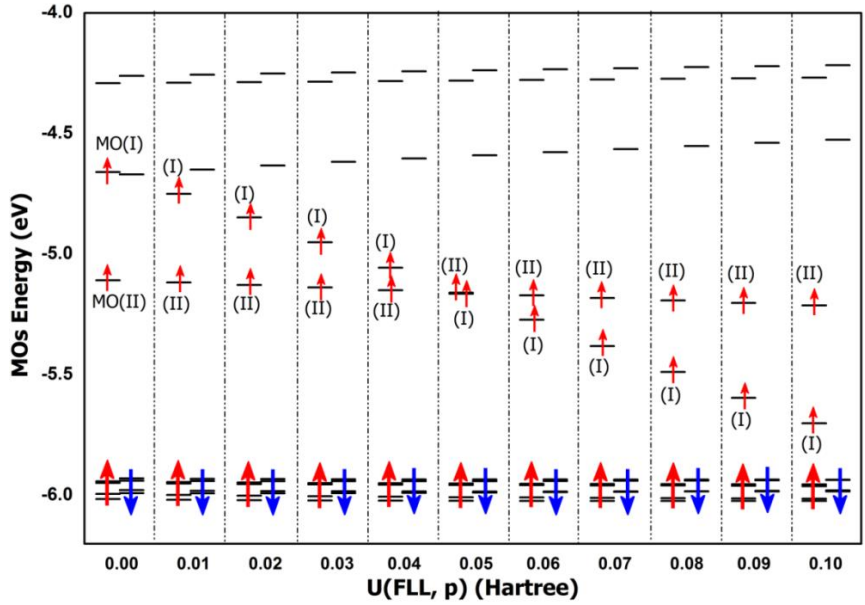

FIG. 3. Electronic structure of $C_{61}$ calculated by the DFTB+U (FLL, p) method.

Next, both the 2s and 2p shells of the carbon atoms are used in the form of the FLL for the U correction, as shown in Table II. The C–C bond length between the adatom and adjacent atoms is slightly elongated with increasing U. When U is 0.10 Ha, the C–C bond length increases by 0.02 Å compared to no U correction. The vertex angle of the defect is significantly reduced, so that the angle with U = 0.10 Ha is smaller than U = 0, by about 3.5 °. For the U (FLL, sp) correction, both the net charge and spin distribution of the adatoms have an increasing tendency with U increase.

TABLE II. Geometric parameters, charge, net spin distribution and HOMO-LUMO gap of $C_{61}$ calculated with the DFTB+U (FLL, sp) method.

| U(FLL, sp) | $l$ | $\angle A$ | Adatom Charge | Adatom Spin | Gap (eV) |
|---|---|---|---|---|---|
| 0.00 | 1.43 | 99.53 | -0.34 | 1.29 | / |
| 0.01 | 1.43 | 99.19 | -0.34 | 1.31 | 0.08 |
| 0.02 | 1.44 | 98.86 | -0.35 | 1.33 | 0.16 |
| 0.03 | 1.44 | 98.53 | -0.35 | 1.33 | 0.25 |
| 0.04 | 1.44 | 98.20 | -0.36 | 1.34 | 0.34 |
| 0.05 | 1.44 | 97.88 | -0.36 | 1.34 | 0.43 |
| 0.06 | 1.44 | 97.55 | -0.37 | 1.35 | 0.52 |
| 0.07 | 1.45 | 97.20 | -0.37 | 1.34 | 0.59 |
| 0.08 | 1.45 | 96.85 | -0.38 | 1.34 | 0.60 |
| 0.09 | 1.45 | 96.47 | -0.39 | 1.33 | 0.61 |
| 0.10 | 1.45 | 96.07 | -0.40 | 1.32 | 0.62 |

The electronic structure of $C_{61}$ calculated by the DFTB+U(FLL, sp) method is shown in Fig. 4. As can be seen, the introduction of U (FLL, sp) can significantly reduce the orbital energy of MO(I) and slightly decrease the orbital energy of MO(II). The influence of U (FLL, sp) on LUMO-β is not monotonous as it increases at first and then decreases, which is the main distinction from the U (FLL, p) method in the computational results. The DFTB+U (FLL, sp) method has no obvious influence on other orbitals. The linear fit of the orbital energies of MO(I) and MO(II) show that they are equal when the U (FLL, sp) is 0.06845 Ha (1.86 eV), as shown in Fig. S2[46]. Meanwhile, when U is less than 0.06 Ha, the data in Table II show that the HOMO-LUMO gap strongly increases with U. The increases in the energy gap are extremely weak when U is larger than 0.06 Ha, similar to the DFTB+U (FLL, p) correction. Thus, the DFTB+U (FLL, sp) method can also reasonably improve the electronic structure and energy gap of $C_{61}$.

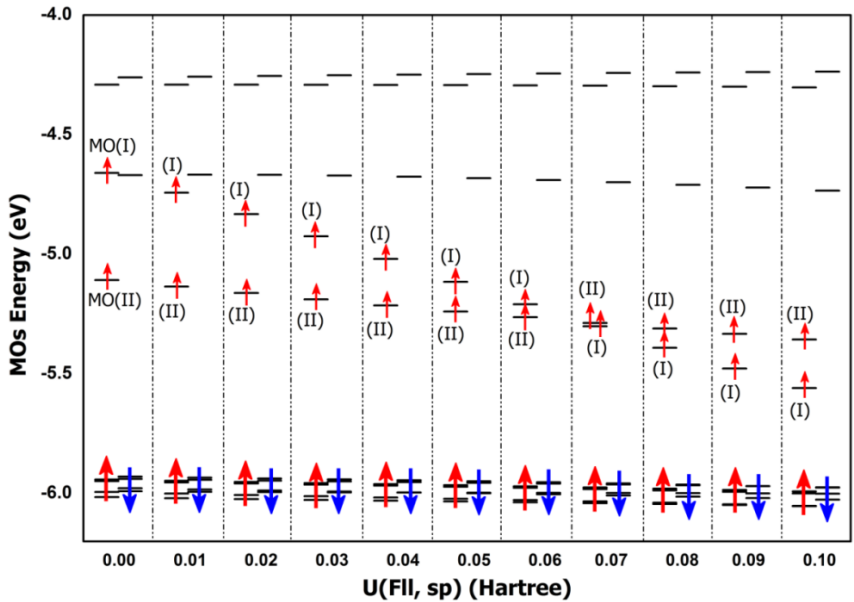

FIG. 4. Electronic structure of $C_{61}$ calculated by the DFTB+U (FLL, sp) method.

After U (FLL, p) and U (FLL, sp) methods, the geometric parameters of $C_{61}$ optimized by the DFTB+U (pSIC, p) method are displayed in Table III. The pSIC potential with respect to the impact of the geometry reflect an opposite trend compared with the FLL potential. The U (pSIC, p) correction scarcely affects the C–C bond length of the defect, but increases the vertex angle of the defect. Compared with U = 0, the vertex angle increases by about 2° when U = 0.10 Ha. With U increasing, the charge of the adatoms has a very slight decreasing trend and the net spin of the adatom gradually increases.

TABLE III. Geometric parameters, charge, net spin distribution and HOMO-LUMO gap of $C_{61}$ calculated with the DFTB+U (pSIC, p) method.

| U(pSIC, p) | $l$ | $\angle A$ | Adatom Charge | Adatom Spin | Gap (eV) |
| --- | --- | --- | --- | --- | --- |
| 0.00 | 1.43 | 99.53 | -0.34 | 1.29 | / |
| 0.01 | 1.43 | 99.70 | -0.34 | 1.32 | 0.13 |
| 0.02 | 1.43 | 99.87 | -0.34 | 1.36 | 0.27 |
| 0.03 | 1.43 | 100.05 | -0.34 | 1.38 | 0.41 |
| 0.04 | 1.43 | 100.23 | -0.34 | 1.40 | 0.55 |
| 0.05 | 1.43 | 100.41 | -0.33 | 1.41 | 0.61 |
| 0.06 | 1.43 | 100.60 | -0.33 | 1.43 | 0.64 |
| 0.07 | 1.42 | 100.78 | -0.33 | 1.45 | 0.67 |
| 0.08 | 1.42 | 100.97 | -0.32 | 1.47 | 0.70 |
| 0.09 | 1.42 | 101.15 | -0.32 | 1.48 | 0.73 |
| 0.10 | 1.42 | 101.34 | -0.31 | 1.51 | 0.76 |

In the calculation of the electronic structure, different from the FLL correction, the pSIC correction can decrease all orbital energies no matter whether they are occupied or unoccupied states, as shown in Fig. 5. However, the orbital energy of MO(I) is the most strongly reduced, meaning that the pSIC correction lowers the orbital energy of the localized occupied state MO(I) and increases the HOMO-LUMO gap. We also linear fit the orbital energies of MO(I) and MO(II) and found the orbital energy of MO(I) is equal to that of MO(II) when U (pSIC, p) is 0.04193 Ha (1.14 eV) (see Fig. S3)[46]. Continuing to increase the value of U leads to an orbital energy of MO(I) which is less than that of MO(II). Thus, the value of U should not be larger than 0.04193 Ha for the U (pSIC, p) correction.

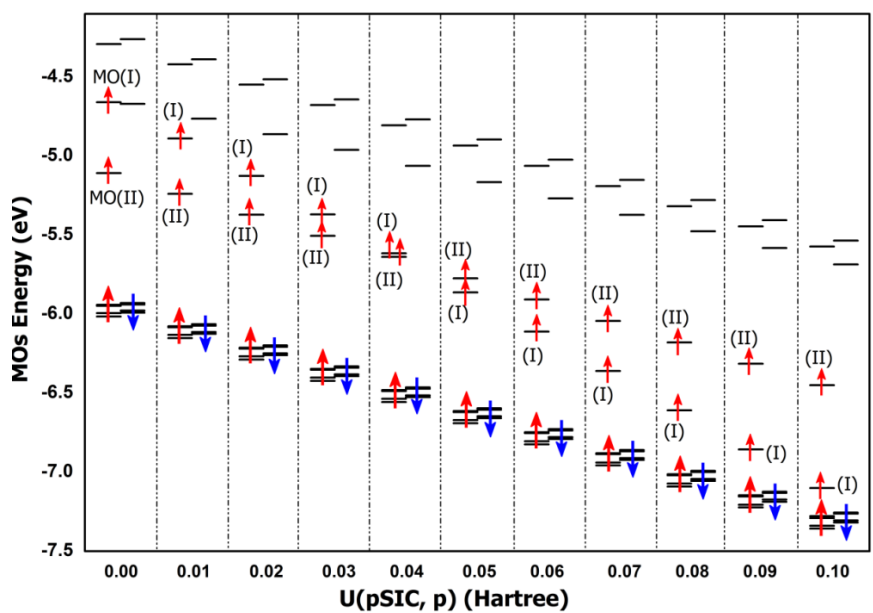

FIG. 5. Electronic structure of $C_{61}$ calculated by the DFTB+U (pSIC, p) method.

Following the former results of three methods, the geometric parameters of $C_{61}$ optimized by the DFTB+U (pSIC, sp) method are displayed in Table IV. The U (pSIC, sp) correction also scarcely affects the C–C bond length, but does decrease the vertex angle of the defect. With U increasing, the charge and net spin of the adatom are slightly increased. For the electronic structure of $C_{61}$, the U (pSIC, sp) correction reflects the features of pSIC shown in Fig. 6. U (pSIC, sp) can also lower all orbital energies and the orbital energy of MO(I) reduces most. Thus, the U (pSIC, sp) correction achieves an effect that the orbital energy of localized occupied states MO(I) relatively reduce and the HOMO-LUMO gap increases. We linear fit the orbital energies of MO(I) and MO(II), and found that the orbital energies of MO(I) and MO(II) are equal when U (pSIC, sp) is 0.06835 Ha (1.86 eV), as shown in Fig. S4[46]. The continuous increase of U contributes to the orbital energy of MO(I) being lower than that of MO(II). Table IV shows that the HOMO-LUMO gap increases with increasing U, similar to the above case. Thus, the U value should be smaller than 0.06835 Ha.

TABLE IV. Geometric parameters, charge, net spin distribution and HOMO-LUMO gap of $C_{61}$ calculated by the DFTB+U (pSIC, sp) method.

| U(pSIC, sp) | $l$ | $\angle A$ | Adatom Charge | Adatom Spin | Gap (eV) |
|---|---|---|---|---|---|
| 0.00 | 1.43 | 99.53 | -0.34 | 1.29 | / |
| 0.01 | 1.43 | 99.19 | -0.34 | 1.31 | 0.08 |
| 0.02 | 1.44 | 98.86 | -0.35 | 1.33 | 0.16 |
| 0.03 | 1.44 | 98.53 | -0.36 | 1.33 | 0.25 |
| 0.04 | 1.44 | 98.20 | -0.36 | 1.34 | 0.34 |
| 0.05 | 1.44 | 97.88 | -0.37 | 1.34 | 0.43 |
| 0.06 | 1.44 | 97.55 | -0.37 | 1.35 | 0.52 |
| 0.07 | 1.45 | 97.20 | -0.37 | 1.34 | 0.59 |
| 0.08 | 1.45 | 96.85 | -0.38 | 1.34 | 0.60 |
| 0.09 | 1.45 | 96.47 | -0.39 | 1.33 | 0.61 |
| 0.10 | 1.45 | 96.07 | -0.40 | 1.32 | 0.62 |

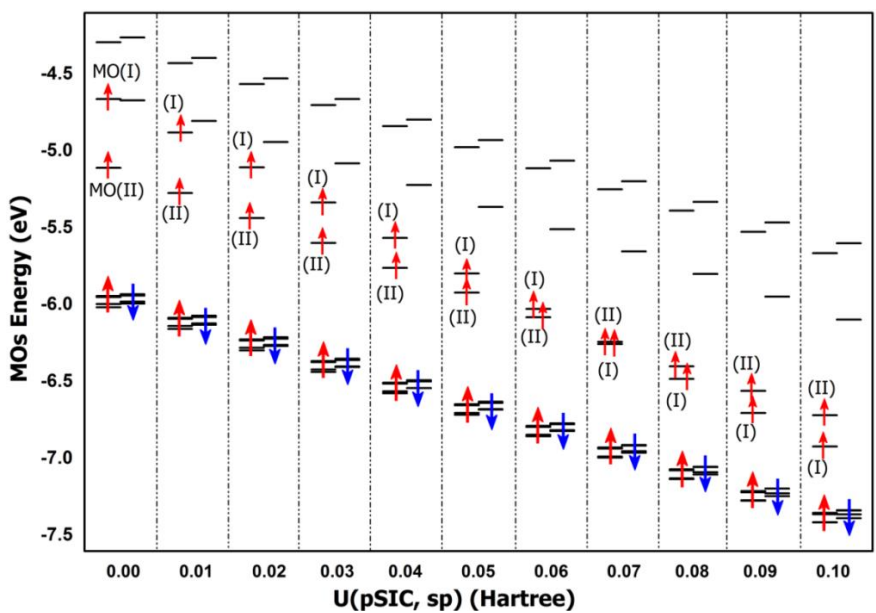

FIG. 6. Electronic structure of $C_{61}$ calculated by the DFTB+U (pSIC, sp) method.

In addition, we summarized the HOMO-LUMO gap calculated by different corrections and the results are shown in Fig. 7. It was found that, for s and p shell corrections, the HOMO-LUMO gaps calculated by the FLL and pSIC corrections are almost the same at U values ranging from 0.01 to 0.10 Ha. For only the p shell correction, the trend in gaps is consistent for the FLL and pSIC corrections calculated, but have slightly differences in values. These results show that the U (FLL, sp) and U (pSIC, sp) corrections have the same function on the HOMO-LUMO gap but U (FLL, p) and U (pSIC, p) have a little difference.

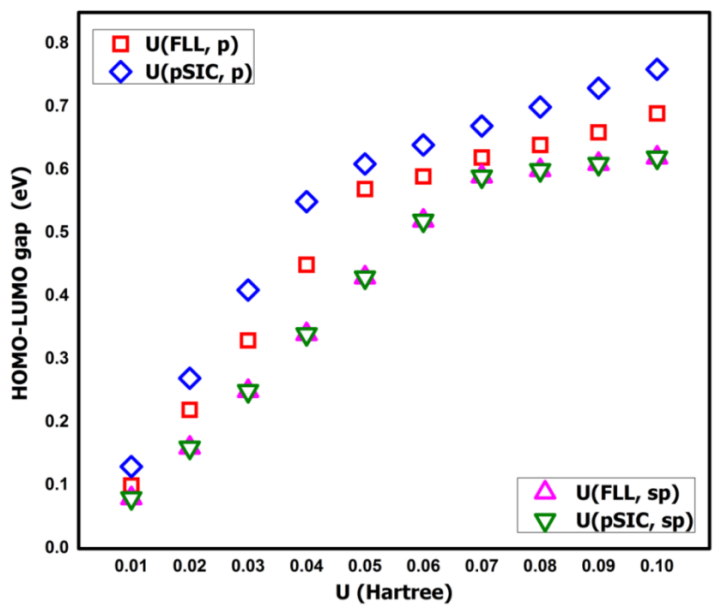

FIG. 7. Change rule of the HOMO-LUMO gap.

## IV. CONCLUSIONS

In summary, we used the DFTB+U method to calculate the electronic structure of a typical carbon adsorbed defect system, namely $C_{61}$. The DFTB calculations show that the adatom has two spin-polarized electrons; this kind of electron behavior is similar to the adsorbed defect behavior of graphene or CNTs, as previously reported. The energy of the lowest unoccupied state, LUMO-β, is even lower than the energy of the highest occupied state, MO(I). However, by using the U correction, the electronic structure and the HOMO-LUMO gap can be significantly improved. For U corrections with the FLL and pSIC, we found that the impact of the U correction on the gap is almost identical, but a larger influence on the geometry is observed when it is added on both the s and p shells. When we only add U correction on the p shell, the two forms of corrections have no obvious effect on the geometric structure, and have almost same influence on the gap. In addition, the FLL correction mainly decreases the orbital energy of MO(I), but the pSIC correction can decrease all orbital energies. In this study, we indicate the dependence of U adjustment on mathematics, and this kind of adjustment needs to be further confirmed by theoretical or experimental research. We hope these results can provide a reference for future research related to the spin polarization of carbon materials based on DFTB methods.

## ACKNOWLEDGMENTS

This work was supported by the National Natural Science Foundation of China (grant number 11374004) and the Science and Technology Development Program of Jilin Province of China (20150519021JH). Z. W. also acknowledges the Fok Ying Tung Education Foundation (142001) and the High Performance Computing Center of Jilin University.